\documentclass{article}
\newcommand{\mv}[1]{\mathbf{#1}}
\newcommand{\fref}[1]{Figure~\ref{#1}}

\usepackage{arxiv}

\usepackage[utf8]{inputenc} 
\usepackage[T1]{fontenc}    
\usepackage{hyperref}       
\usepackage{url}            
\usepackage{booktabs}       
\usepackage{amsfonts}       
\usepackage{nicefrac}       
\usepackage{microtype}      
\usepackage{graphicx}
\usepackage{amsmath}
\usepackage{cite}

\title{Hybrid real- and reciprocal-space full-field imaging with coherent illumination}

\author{
  Po-Nan Li\\
  Department of Electrical Engineering\\
  Stanford University\\
  Stanford, CA 94305, USA\\
  \texttt{liponan@stanford.edu} \\
   \And
  Soichi Wakatsuki\\
  Department of Structural Biology\\
  Stanford University\\
  Stanford, CA 90305, USA\\
  \texttt{soichi.wakatsuki@stanford.edu} \\
  \And
  Piero A. Pianetta\\
  Stanford Synchrotron Radiation Lightsource\\
  SLAC National Accelerator Laboratory\\
  Menlo Park, CA 94025, USA\\
  \texttt{pianetta@slac.stanford.edu} \\
  \And
  Yijin Liu\\
  Stanford Synchrotron Radiation Lightsource\\
  SLAC National Accelerator Laboratory\\
  Menlo Park, CA 94025, USA\\
  \texttt{liuyijin@slac.stanford.edu} \\
}

\begin{document}
\maketitle

\begin{abstract}
We present a novel diffractive imaging method that harnesses a low-resolution real-space image to guide the phase retrieval. A computational algorithm is developed to utilize such prior knowledge as a real-space constraint in the iterative phase retrieval procedure. Numerical simulations and proof-of-concept experiments are carried out, demonstrating our method's capability of reconstructing high-resolution details that are otherwise inaccessible with traditional phasing algorithms. With the present method, we formulate a conceptual design for the coherent imaging experiments at a next-generation X-ray light source.
\end{abstract}

\section{Introduction}
Phase retrieval, the procedure in which the phase is recovered from the measured intensity pattern(s), is crucial for diffraction-based imaging approaches and has been extensively studied in the past decades.
In Fienup's pioneering work, the hybrid input-output (HIO) algorithm was presented and is widely used by researchers in this field \cite{Fienup1982, Bauschke2002, Fienup2012}.
HIO not only exhibits an elegant mathematical formulation, but also demonstrates superior versatility and robustness against the stagnation problem, which is often a significant issue when utilizing the other precedent algorithms.
HIO starts with inversely Fourier transforming the observed magnitude profile in the reciprocal-space (supplemented with a set of random phases as the initial guess) to the real-space, where some constraints are applied to update the real-space image.
A forward Fourier transform is then executed to bring the data back to the reciprocal-space, where the Fourier modulus is replaced by the measured value and the calculated phase is kept.
Such consecutive inverse and forward Fourier transformations are repeated until it reaches to a convergence or until a meaningful real-space image is achieved. 

X-ray coherent diffraction imaging (XCDI), by which the sample is illuminated by a coherent X-ray light source and the resulting far-field diffraction pattern is phased to recover the image, has been regarded as a potentially groundbreaking development in X-ray microscopy and as one of the major motivations for the development of X-ray free electron lasers (X-FEL).
XCDI relies on the numerical algorithms like HIO to execute the phase retrieval, which yields the real-space images \cite{Sayre1998, Miao1998, Miao1999, Miao2012}.
To this end, several alternatives and modifications to the original HIO have been proposed and widely studied, such as shrink-wrap \cite{Chapman2006}, GHIO \cite{Chen2007} and OSS \cite{Rodriguez2013}.
In addition to the algorithm developments, modified experimental schemes for XCDI have been proposed for achieving better spatial resolution and improved robustness of phase retrieval.
Well-known examples include key-hole holography \cite{Abbey2008}, in-flight holography \cite{Gorkhover2018} and the approaches using a metallic template \cite{Lan2014} or a block-reference \cite{Barmherzig2019}. 

Despite all the advancements, phase retrieval in XCDI is still one of the major obstacles that hinders the broad application of XCDI.
It is nontrivial to construct a robust phase retrieval workflow that eliminates the vulnerability to common sources of image imperfection, e.g. missing data and quantum noise.
Missing low-$q$ data in the diffraction pattern is often caused by a beam stop that blocks the direct beam to protect the detector.
Quantum noise is inevitable in diffractive imaging and it largely depends on the brilliance of the light source. 
Although these adversities could potentially be alleviated by the projected major upgrades of the X-ray facilities and beamlines, here we demonstrate an alternative approach that is effective and practical. 

In this work, we demonstrate a hybrid approach, in which we image the same sample with two different modalities: a coherent diffraction imaging apparatus provides a diffraction pattern and a direct imaging setup gives a real-space image, where the latter intrinsically has a lower image resolution due to the limited numerical aperture of the objective lens.
We develop a phase retrieval algorithm that uses the low-resolution (LR) image as a real space constraint and reconstructs high resolution details from a diffraction pattern recorded from the same object.
The contribution of this paper is three-fold: (i) we provide a phase retrieval algorithm that takes advantage of the LR image; (ii) we provide an experimental proof-of-concept with a He-Ne laser-based table-top setup; and (iii) we conceive a novel conceptual design of an experimental setup that allows the collection of diffraction patterns and LR images simultaneously at an X-FEL beamline.

\section{Principle}
The present approach uses an LR real-space image of the sample as \textit{a priori} information to guide the high-resolution image reconstruction.
Such LR images can be obtained from a variety of sources and can be acquired concurrently or consecutively, experimentally or computationally.
To take advantage of such LR images for the phase retrieval, we have developed a modified phase retrieval algorithm, which works as follows.
A random phase component is assigned to the square-root of the experimentally measured diffraction intensity, making the input $F^{(0)}(\mv{k})$. 
The input is then inversely Fourier-transformed to the real-space, where two constraints are applied.
In addition to the conventional support constraint, which is based on the estimated shape of the non-zero region and the non-negativity, here we also enforce an LR constraint.
The modified real-space reads
\begin{equation}
    \rho'(\mv{r}) = 
    \begin{cases}
        \rho^{(t)}(\mv{r}) + \gamma \left[ g(\mv{r}) - \rho^{(t)}(\mv{r}) \right]& \mv{r} \in \mv{S} \\
        \rho^{(t-1)}(\mv{r}) - \beta \rho^{(t)}(\mv{r}) & \mv{r} \notin \mv{S}
    \end{cases}
\end{equation}
where $\rho^{(t)}(\mv{r}))$ is the inverse Fourier transform of $F'^{(t-1)}(\mv{k})$, $g(\mv{r})$ is the LR prior and $\mv{S}$ is the real space support.
The feedback parameter $\beta$ is typically set to $0.9$ as discussed in Ref. \cite{Fienup1982}.
The basic idea behind this approach is that there shall be a consistency between the input real-space LR prior and the low-spatial-frequency component of the final high-resolution image.
This consistency is, subsequently, utilized as an additional constraint that guides the update of the real-space image throughout the iterations.

\begin{figure}[htbp]
\centering\includegraphics[scale=1]{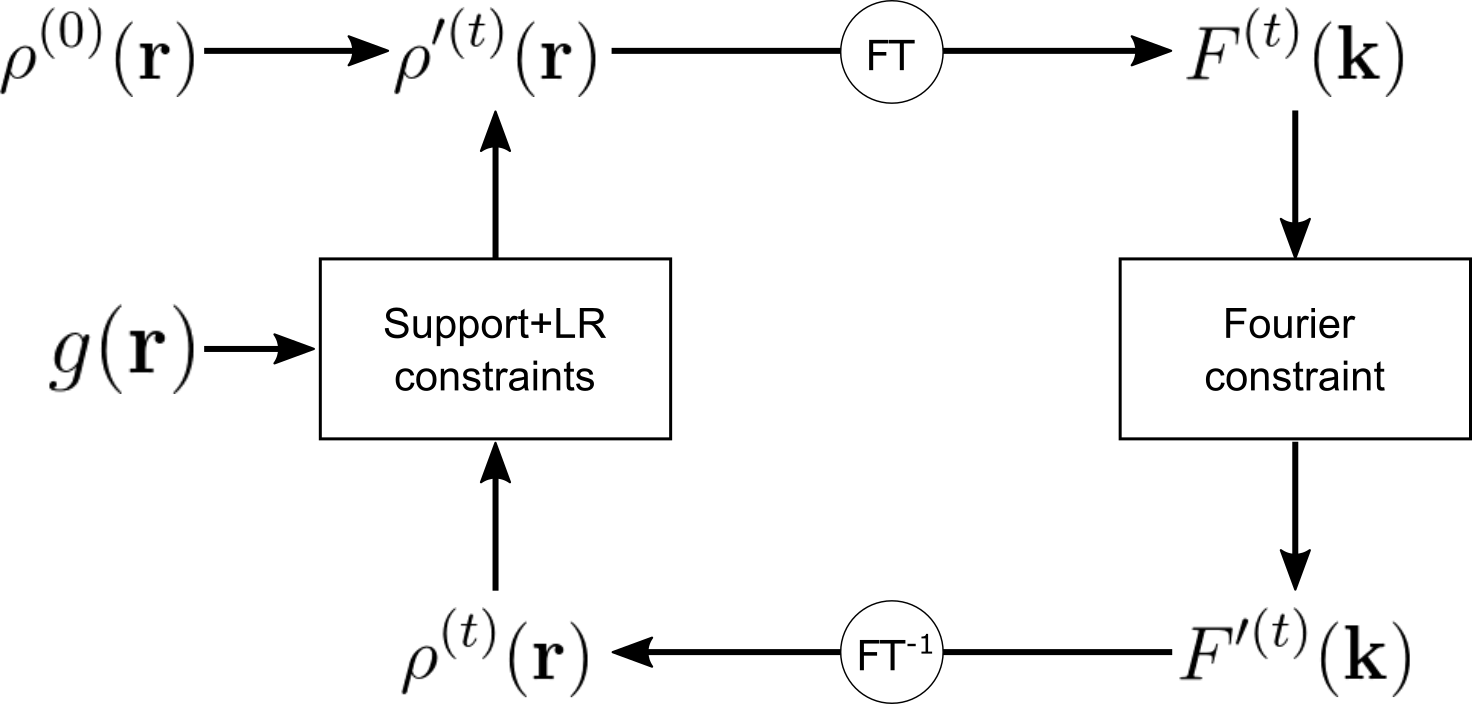}
\caption{\label{fig:algorithm}Schematic illustration of the proposed LR image-guided iterative phase retrieval algorithm. An LR image, denoted $g(\mv{r})$, is used as an additional constraint for the real-space. Input $\rho^{(0)}(\mv{r})$ is given by the inverse Fourier transform of the magnitude of the diffraction pattern $|F^{(0)}(\mv{k})|$ times the phase component with initial random phase $\phi^{(0)}$.}
\end{figure}

To execute the idea discussed above, we introduce a new parameter $\gamma$ that determines the strength to which the LR constraint is enforced.
This parameter should be adaptive.
A strong enforcement is desired during the early iterations for a strong guidance using the LR.
This LR constraint, however, should eventually vanish as one would anticipate achieving high resolution in the final result with high-spatial-frequency information that is not available in the LR.
Intuitively, we can connect $\gamma$ to a figure of merit (FOM) that evaluates the image quality on-the-fly.
One example is the normalized Fourier space residual function
\begin{equation}
    \gamma_{F} = \frac{\sum_{\mv{k} \not\in \mv{U}} \left|\left|F(\mv{k})\right| - \left|F(\mv{k})^{\mathrm{(obs)}}\right|\right|}
    {\sum_{\mv{k} \not\in \mv{U}} |F(\mv{k})^{\mathrm{(obs)}}|}
\end{equation}
where $\mv{U}$ denotes the missing data region.
As one can see, if the calculated Fourier modulus is consistent with the measured amplitude, $\gamma_F$ will be close to zero.
The other commonly used FOM is
\begin{equation}
    \gamma_0 = 
    \frac{\sum_{\mv{r} \notin \mv{S}} | \rho(\mv{r}) |^2}
    {\sum_{\mv{r}} |\rho(\mv{r})|^2}
\end{equation}
which computes the ``energy'' outside of the support, as discussed in \cite{Fienup1982}.
Either way, $\gamma$ is directly linked with the FOM, offering an effective on-the-fly image quality assessment.
Later we will show that either FOM function will suffice.
In the rest of this paper, the value of $\gamma$ will be determined at each step by a FOM of choice, i.e., either $\gamma = \gamma_F$ or $\gamma = \gamma_0$.
After the real space constraints are applied, the Fourier magnitude constraint follows to ensure consistency with the experimental observation:
\begin{equation}
    F'^{(t)}(\mv{k}) =
    \begin{cases}
        |F^{(t)}(\mv{k})|\exp j \phi^{(t)}, & \mv{k} \in \mv{U}\\
        |F^{(0)}(\mv{k})|\exp j \phi^{(t)}, & \mv{k} \notin \mv{U}
    \end{cases}
\end{equation}
where $|F^{(0)}(\mv{k})|$ is the square-root of the experimentally measured diffraction intensity, $|F^{(t)}(\mv{k})|$ and $\phi^{(t)}$ are the magnitude and phase of the Fourier transform of $\rho'(\mv{r})$, respectively, and $\mv{U}$ denotes the region with missing data.
For a more intuitive presentation, we depict in \fref{fig:algorithm} the workflow of the herein developed LR-guided phase retrieval algorithm.

\section{Simulation}
\begin{figure}[htbp]
\centering\includegraphics[scale=1]{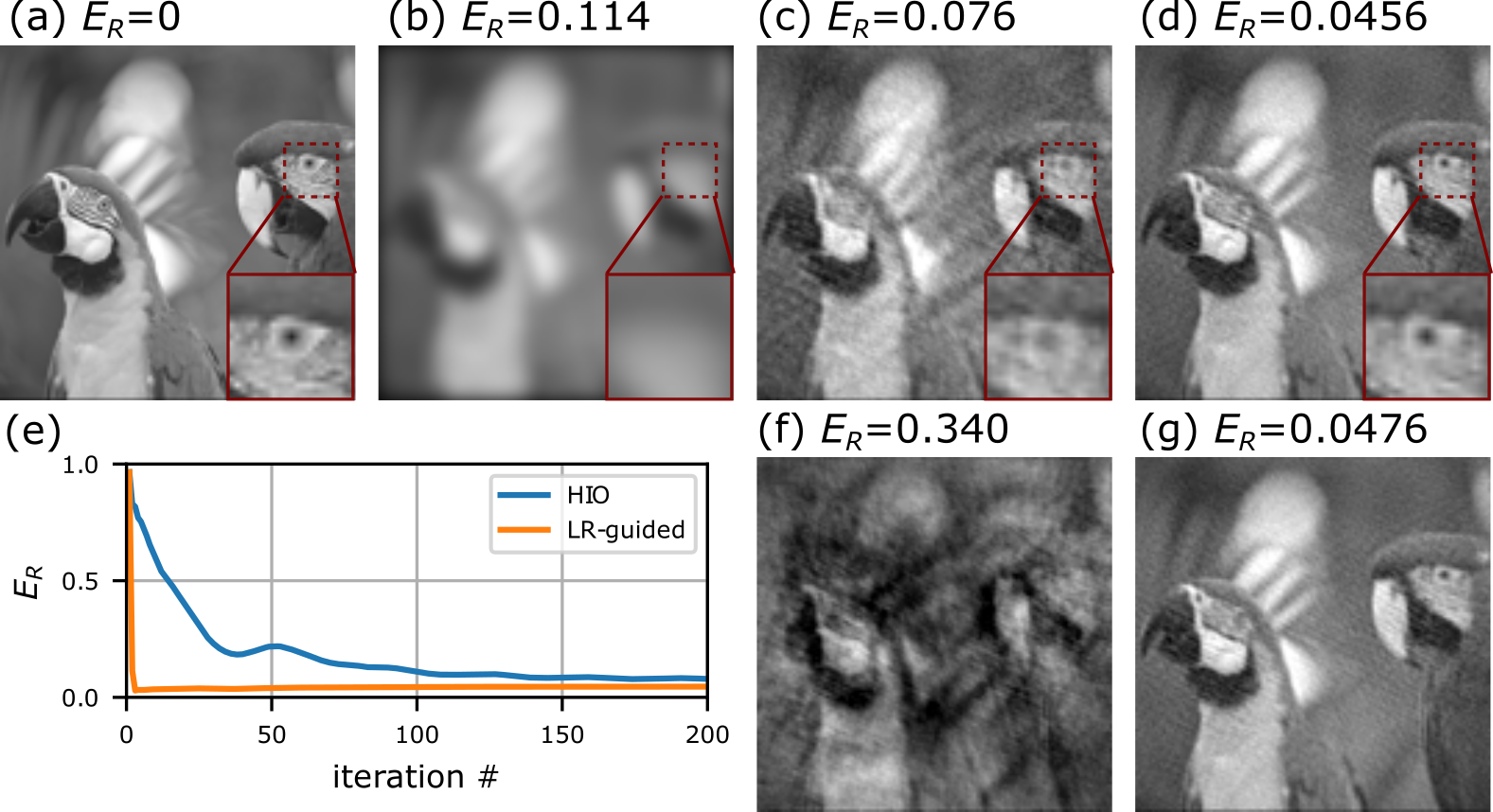}
\caption{\label{fig:simulation}Simulation results. (a) Ground truth. (b) The LR image. (c) Reconstruction with HIO, $\eta=1$. (d) Reconstruction with our LR-guided method, with $\eta=1$. Inset in each panel shows the zoom-in of the respective highlighted region. (e) Real space residual ($E_R$) vs. iteration number. (f, g) Reconstruction results with HIO and our method, respectively, with $\eta = 3$.}
\end{figure}

We cropped an image (``Two macaws'') from the Kodak image set and resampled to $128 \times 128$ pixels (px) (\fref{fig:simulation}(a)), then zero-padded to $640 \times 640$ pixels to simulate $5\times$ oversampling along each dimension.
We assume the photon density of $5 \times 10^{7}$/px at the center of the diffraction pattern and use Poisson statistics to simulate the quantum noise. 
$11 \times 11$ pixels at the center were removed to mimic the missing center with $\eta = 1$, i.e. missing $1$ wave component, as defined by Miao \textit{et al.} \cite{Miao2005}.
An exact support was used in our simulations.
The LR image, shown in \fref{fig:simulation}(b), was generated by applying a Gaussian mask $H(\mv{k})=\exp\left(-\mv{k}^2/2\sigma^2\right)$ in the reciprocal-space, where $k \in (-0.5, 0.5)$ is the spatial frequency in px$^{-1}$, with bandwidth factor $\sigma = 0.05$ px$^{-1}$ along each dimension.
Each phase retrieval algorithm started from the same random phase set and ran for $200$ iterations.

The reconstruction result using our method (\fref{fig:simulation}(d)) is clearly better than that of the HIO (\fref{fig:simulation}(c)) by means of direct visual assessment.
To further quantify the improvement, we use the real-space R-factor function
\begin{equation}
    E_R = 
    \frac{\sum_{\mv{r} \in \mv{S}} |\rho(\mv{r}) - \rho(\mv{r})^{(\mathrm{model})}|}{ \sum_{\mv{r} \in \mv{S}} \rho(\mv{r})^{(\mathrm{model})}}
\end{equation}
to evaluate the quality of image $\rho(\mv{r})$ in real-space against the model $\rho(\mv{r})^{(\mathrm{model})}$.
\fref{fig:simulation}(e) shows that our method not only converges faster than HIO, but also achieves considerably improved image quality as evidenced by the suppressed $E_R$ value. 

It is of practical importance to evaluate the scenario of more missing data in the center of the diffraction pattern, which is a common issue in this field and could be restricted by a number of practical constraints.
For this purpose, we further increased the area of the missing center, from $\eta = 1$ to $\eta = 3$, and conducted the phase retrieval under this condition.
The HIO result, as shown in \fref{fig:simulation}(f), failed to faithfully represent the ground truth; whereas our result, as shown in \fref{fig:simulation}(g), clearly demonstrates superior quality and the $E_R$ value stays almost the same as the case of $\eta = 1$.
Note that the artifacts in \fref{fig:simulation}(d) and (g) are due to the quantum noise added to the diffraction pattern.

\begin{figure}[htbp]
\centering\includegraphics[scale=1]{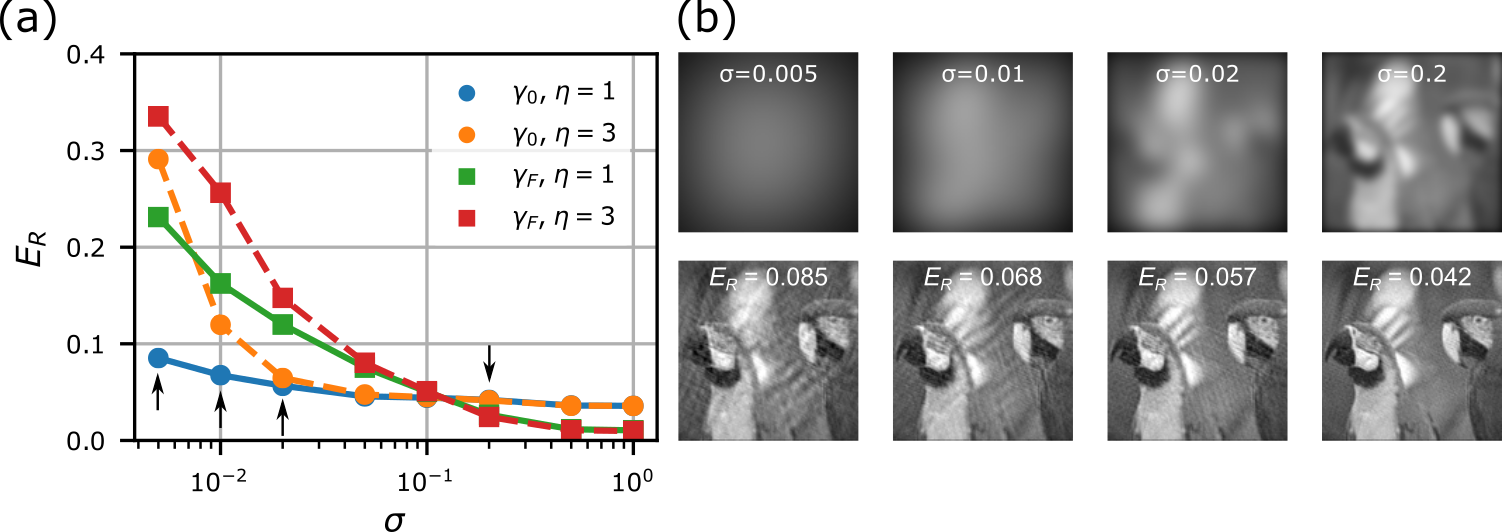}
\caption{\label{fig:er_vs_sig}(a) Reconstruction error with various conditions: choice of FOM function, number of missing waves $\eta$ and bandwidth factor $\sigma$ of the LR image . Arrows indicate data points to be displayed in (b). (b) Selected pairs of LR images with various values of the bandwidth factor $\sigma$ (upper row, unit: px$^{-1}$) and their corresponding reconstructed images by our method with $\gamma_0$ and $\eta = 1$ (lower row).}
\end{figure}

Intuitively, one would expect that the better resolution in the LR prior, the more information it can provide and, thus, the more likely it will render a good reconstruction.
Our result in \fref{fig:er_vs_sig}(a) clearly supports this.
We tested our algorithm using a number of different LR images with different spatial resolutions as the real-space constraint (\fref{fig:er_vs_sig}(f)).
Strikingly, \fref{fig:er_vs_sig}(a) reveals that even a dramatically blurred LR image ($\sigma=0.005$ px$^{-1}$, top far left in \fref{fig:er_vs_sig}(b)) still facilitates the LR-guided approach to deliver an excellent reconstruction (with $E_R=0.23$ at $\eta=3$ and $\gamma_0$ is used as the FOM function).
Furthermore, \fref{fig:er_vs_sig}(a) also suggests that using an LR image with $\sigma=0.02$, meaning with a resolution at $8.5\times$ coarser ($3$dB cutoff) than the ultimate target, will suffice to enable a very robust reconstruction in the scenario with significant missing data region ($\eta=3$).

\section{Experiments and results}
\begin{figure}[htbp]
\centering\includegraphics[scale=0.6]{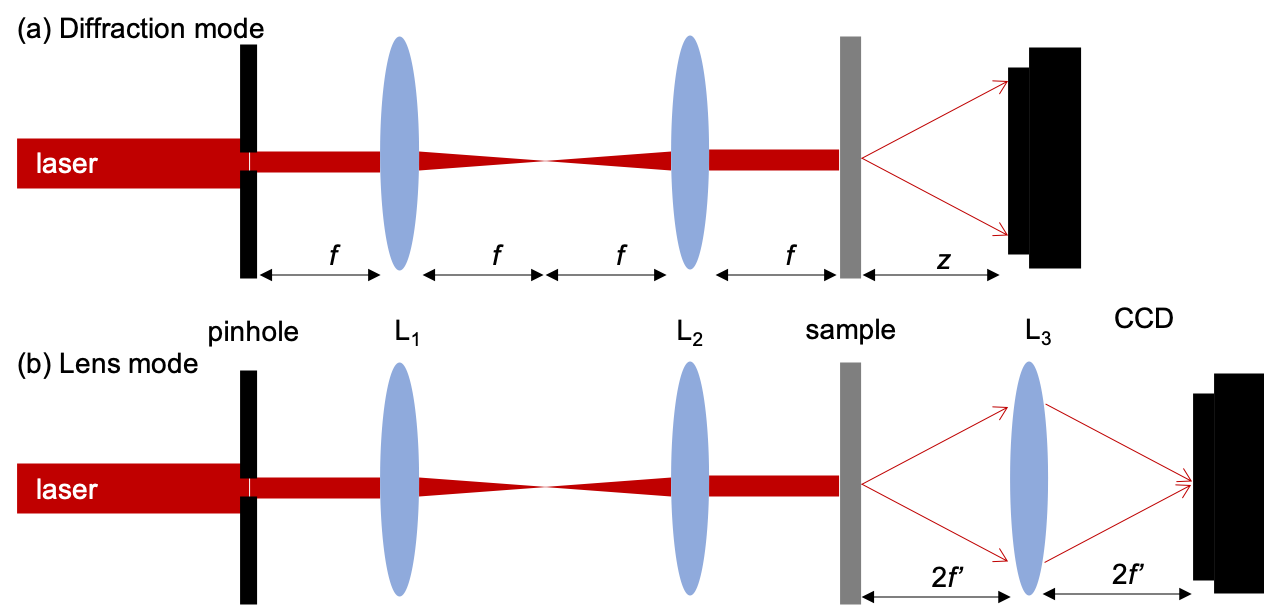}
\caption{\label{fig:table_top}Schematic illustration of the table-top optical setup we constructed for the proof-of-concept experiment in (a) the diffraction mode and (b) the lens mode. $f$ and $f'$ are local lengths of lenses $L_i$; $z$ is sample-to-camera distance.}
\end{figure}

To provide an experimental proof-of-concept, we performed table-top optical experiments, in which we acquired diffraction and LR images in a serial manner.
We used a setup with two bi-convex lenses to relay the image of a $200$ $\mu$m diameter pinhole onto the sample with unity magnification, as shown in \fref{fig:table_top}(a).
We imaged one of the pattern groups on a 1951 USAF resolution target (Thorlabs) as described below.
An air-cooled CCD camera (Starlight Xpress SXVR-H694) was placed approximately $32$ mm downstream of the sample to record diffraction patterns.
We took $10$ images with each of the following exposure times: $100$, $1000$ and $10000$ ms.
Averaged images of each time exposure group were subtracted with a dark image and then assembled to a high dynamic range diffraction image of $2201\times2201$ pixels.
After we had done the measurement of the diffraction patterns, a bi-convex was then inserted between the sample and the detector, as shown in \fref{fig:table_top}(b), to capture LR images.
The LR image is the average of $10$ images with exposure time of $1$ ms and subtracted with the same dark image.

\begin{figure}[htbp]
\centering\includegraphics[scale=1.0]{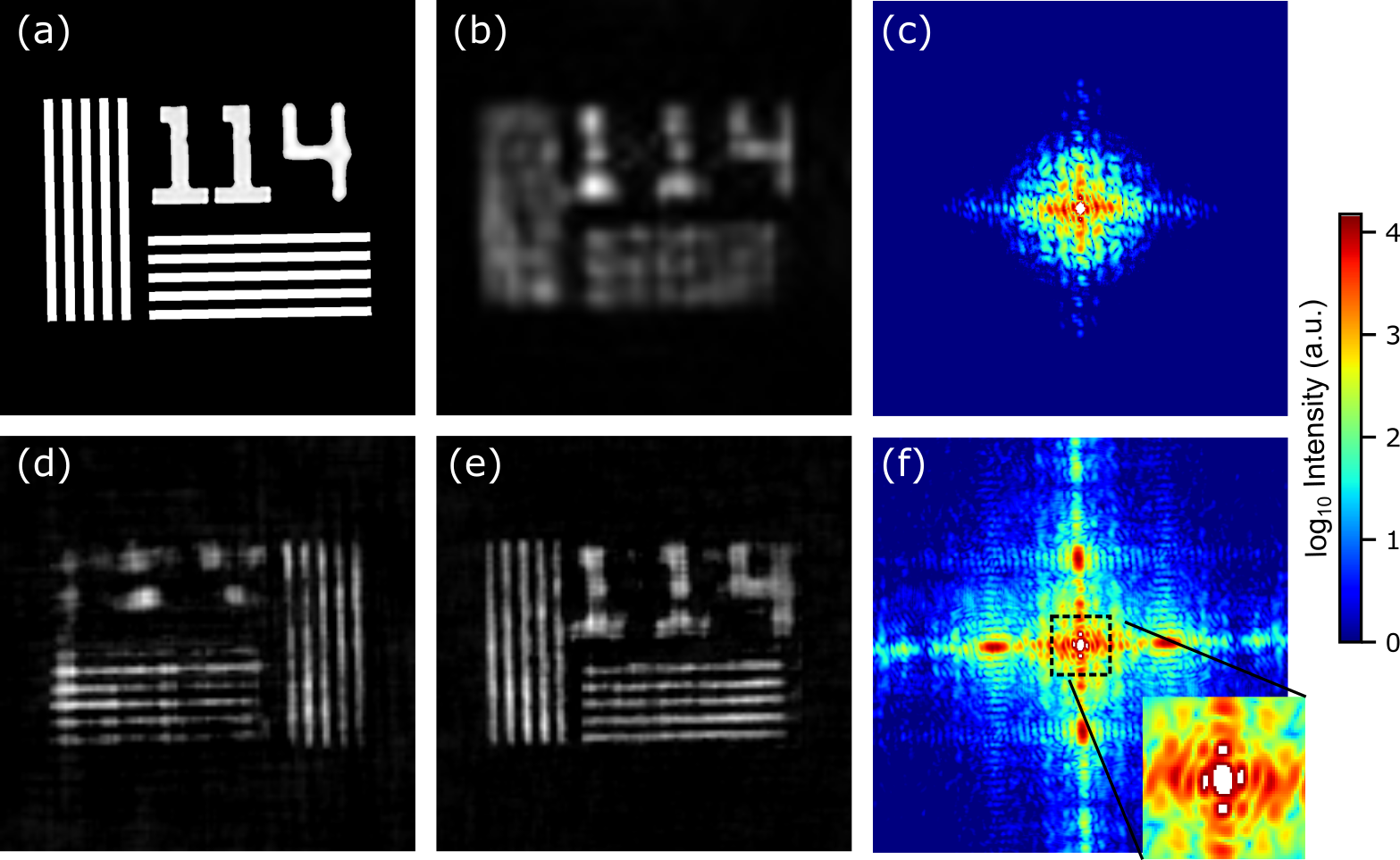}
\caption{\label{fig:tabletop_results}(a) Illustration of the group 114 on a 1951 USAF resolution target. (b) The LR image measured using the diffraction mode depicted in \fref{fig:table_top}(b).  (c) Fourier transform of (b) (logarithmic scale).  (d) Reconstruction using the HIO algorithm. (e) Reconstruction result of our algorithm. (f) Experimentally measured diffraction pattern (logarithmic scale). Inset: zoom-in of the center area, where the saturated (white) region is treated as missing data during the phase retrieval.}
\end{figure}

We imaged the $114$ group on the 1951 USAF resolution target, as shown in \fref{fig:tabletop_results}(a).
The number $114$ indicates the arrangement of the vertical and horizontal lines is $114$ cycles per mm, equivalent to $8.77$ $\mu$m per cycle. \fref{fig:tabletop_results}(b) and (c) show the LR image we collected with the lens mode depicted in \fref{fig:table_top}(b) and its Fourier transform, respectively.
We conducted the phase retrieval using the pattern we collected in the diffraction mode (\fref{fig:table_top}(a)) with HIO and our LR-guided algorithm.
\fref{fig:tabletop_results}(d) and (e) show unambiguously that our reconstruction result (e) outperforms the HIO result (d) and is capable of correctly resolving the fine horizontal and vertical lines, which are hardly visible in the LR image (\fref{fig:tabletop_results}(b)).
In contrast, the HIO result is not recognizable, in addition, it has the horizontal and vertical lines in a highly distorted arrangement. 
It is also clear, by comparing \fref{fig:tabletop_results}(c) and (f), the Fourier transform of the LR image and the diffraction pattern, that the LR image only has information within a relatively narrow bandwidth.
We point out here that in the experiments, we purposely saturated the detector to mimic the realistic experimental conditions, in which the low-$q$ data is not accessible, as shown in \fref{fig:tabletop_results}(f) inset.

\section{A proposed X-FEL experimental setup}
\begin{figure}[htbp]
\centering\includegraphics[scale=1]{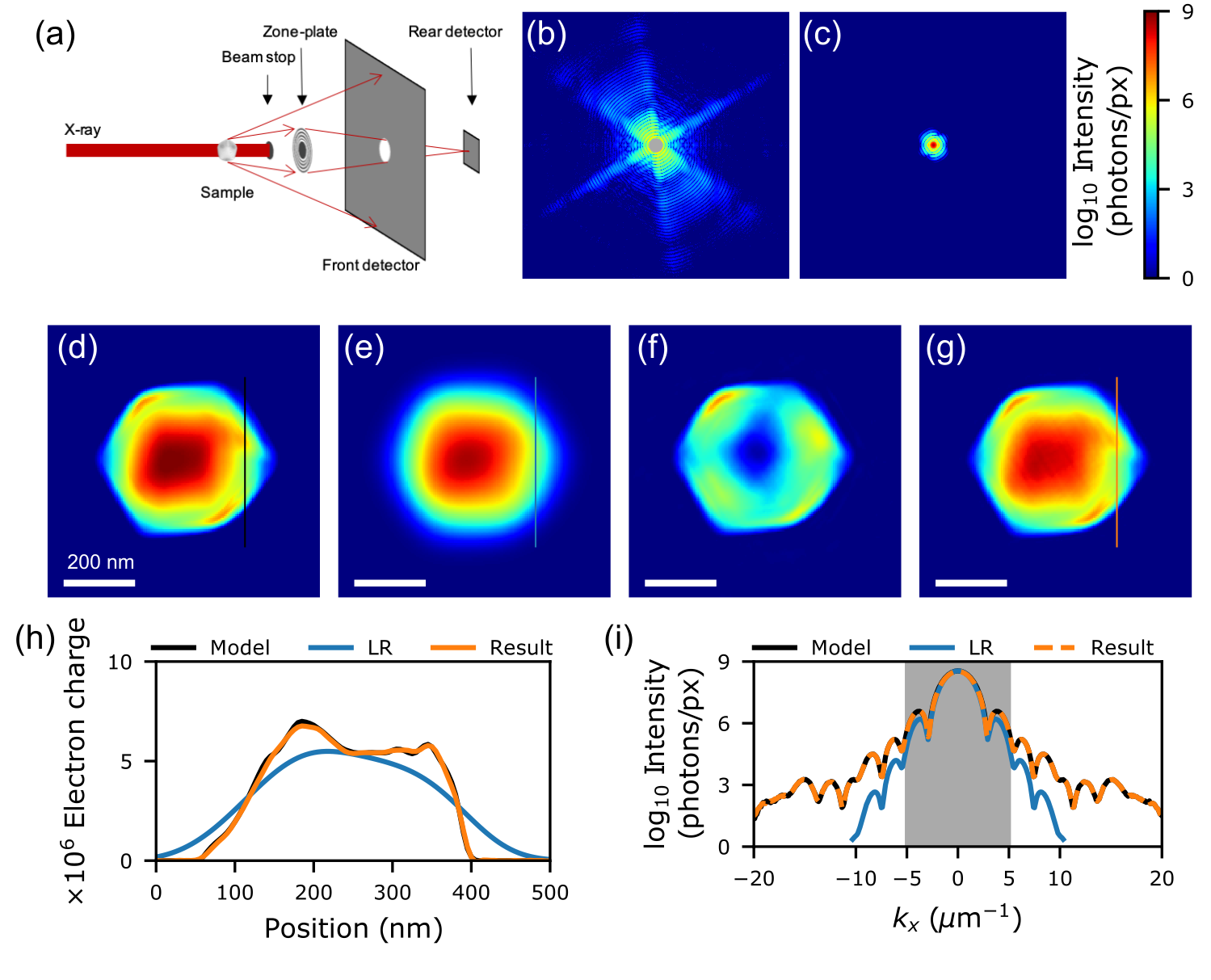}
\caption{\label{fig:xfel_design}Simulated Mimivirus hybrid imaging experiment. (a) Proposed imaging set-up (not drawn to scale).  (b) Simulated diffraction pattern of the mimivirus at the front detector in (a). The shaded circle marks the missing data area. (c) Fourier intensity of the LR image, i.e., (e). (d) Model derived from the projection of the published EM data \cite{Xiao2009}. (e) LR image, as if the real image recorded by the rear detector in (a). (f) Result with HIO. (g) Reconstruction with our method, starting from the same random phase as that used for HIO. (h) Comparison of line profiles in the model (d), the LR image (e), and our reconstruction result (g). (i) Comparison of Fourier spectra at $k_y=0$. The shaded region indicates the missing data area corresponding to the circle in (b).}
\end{figure}

Our demonstrations in the numerical simulation and the proof-of-concept laser experiment are encouraging.
The present imaging method is expected to be directly adaptable in X-ray imaging experiments at X-ray light source facilities.
To further explore such a possibility, we simulate a Mimivirus imaging experiment in an X-FEL setup \cite{Ekeberg2016}.
\fref{fig:xfel_design}(a) shows a schematic of the proposed diffraction imaging setup, where a focusing element (e.g. a Fresnel zone-plate) is placed in front of the first detector and is used to capture the low-$q$ component that would otherwise be dumped or blocked.
This zone plate serves as an objective lens that records the low-resolution real-space image concurrent with the acquisition of the X-ray diffraction pattern.
We assume an X-ray wavelength of $1.03$ nm ($1.2$ keV) and the distance between the sample and the front of the detector to be $1.65$ m so the full-period resolution at the edge of the front detector is $10$ nm.
The zone-plate images the sample onto the rear detector \cite{Tkachuk2007, Mao2019}.
\fref{fig:xfel_design}(b) shows the simulated diffraction pattern that would be recorded by the front detector, e.g. a Rayonix MX340-XFEL detector \cite{Sierra2019}, one of the detectors utilized at LCLS that has $1920\times1920$ pixels with pixel size of $177$ $\mu$m ($4\times4$ binning mode) and a circular hole at its center to allow the intense X-ray beam to pass through.
The shaded region at the center in \fref{fig:xfel_design}(b) is where diffraction patterns are not recorded because of the circular hole.
The electron density profile is derived from the published Cryo-EM map data (EMD-5039) \cite{Xiao2009}.
We assume the incoming photon flux density $2\times10^{12}$ photons/$\mu$m$^{2}$/pulse and use Poisson statistics to simulate the quantum noise.
\fref{fig:xfel_design}(e) and (c) show the simulated image that would be collected by the rear detector and its Fourier transform, respectively.
To simulate limited resolution caused by the limited numerical aperture and the radius of the outermost ring of the zone plate (as shown in \fref{fig:xfel_design}(a)), we low-pass filter the ground-truth image in the Fourier space with $\sigma=0.02$ px$^{-1}$ (equivalent to a point spread function in real-space with full width at half maximum of $65$ nm), then perform an inverse Fourier transform to obtain the real-space image as shown in \fref{fig:xfel_design}(e).
We use a loose support and fixed set of random phases for phase retrieval.
\fref{fig:xfel_design}(f) and (g) show the reconstruction results by using HIO and our algorithm, respectively.
Our result, with error $E_R=0.0345$, not only outperforms the traditional method, but also retrieves details that are apparently not available in the LR image ($E_R=0.147$, \fref{fig:xfel_design}(e)).
To further highlight the capability of our method in recovering the missing low-$q$ data, we compare the line profiles of the low-resolution image and our reconstruction result (\fref{fig:xfel_design}(h)).
Our algorithm is capable of recovering the Fourier intensity in the missing center by incorporating the low-resolution image and reconstructing high resolution details, as shown in \fref{fig:xfel_design}(i).

\section{Discussion}
Our method makes use of an LR real-space image to deliver a successful image reconstruction.
At one extreme, if the \textit{a priori} information barely provides the shape of the sample, it is similar to running the HIO with a tight support or an advanced phase retrieval capable of generating an estimated support.
On the other hand, if the \textit{a priori} information is as good as the ground truth would be, then the algorithm would converge in a few steps without providing much improvement.
Our algorithm leverages an FOM function to assess the reconstruction quality on-the-fly and, thus, can adaptively control the strength that the LR constraint is applied.
As we demonstrated in \fref{fig:er_vs_sig}(a), $\gamma_F$ tends to work very well when $\sigma$ is close to $1$, i.e., the LR image has a relatively high resolution, whereas $\gamma_0$ achieves a very low $E_R$ even when $\sigma$ is small, which is more likely to be the case in real-world applications.
In this work, we use $\sigma$, the bandwidth parameter in an attempt to provide a way to evaluate the relative resolution against the diffraction pattern.
While we realize that it might be difficult to experimentally determine a $\sigma$ value for each image, in general, we propose using $\gamma_0$ as it is robust to an LR with significantly coarsened resolution.

It is useful to mention that the pioneers of XCDI have used the Fourier transform of a low resolution image as a patch to circumvent the missing center \cite{Miao2003}, and later, other researchers have also learned that the missing intensities at the beam stop region can be set as variables and solved by the phase retrieval algorithms \cite{Miao2005}.
Additionally, electron microscopists have also used the Fourier transform of a low-resolution image to enhance the electron diffraction images \cite{DeCaro2010}.
Unlike other work, here we use the LR image in real-space and use it throughout the iterations to guide the phase retrieval, not just as the initial guess.
The advantage of our approach is two-fold: leveraging the low-resolution image itself, rather than its Fourier transform, largely reduces the burden of stitching and rescaling in the reciprocal space; our adaptive constraint ensures that the \textit{a priori} information will not be washed away until the algorithm has significantly converged.

\section{Conclusion}
In this paper we demonstrated a novel coherent diffraction imaging approach that takes advantage of a real-space low-resolution image as the \textit{a priori} information to enhance the quality and robustness of the phase retrieval.
Simulation and proof-of-concept experimental results show that the reconstructed images by our approach are superior than the results by the conventional phasing method.
The present approach could potentially find applications at synchrotron radiation and X-FEL beamlines, providing a robust, high-resolution diffraction imaging solution.

\section*{Funding}
U.S. Department of Energy (DOE), Office of Science, Office of Basic Energy Sciences (DE-AC02-76SF00515).

\section*{Acknowledgements}
The authors thank Jeff Corbett for help and insightful discussions.

\bibliographystyle{unsrt}  

\bibliography{mendeley}

\begin{thebibliography}{10}

\bibitem{Fienup1982}
J.~R. Fienup.
\newblock {Phase retrieval algorithms: a comparison}.
\newblock {\em Applied Optics}, 21(15):2758, aug 1982.

\bibitem{Bauschke2002}
Heinz~H Bauschke, Patrick~L Combettes, and D~Russell Luke.
\newblock {Phase retrieval, error reduction algorithm, and Fienup variants: a
  view from convex optimization}.
\newblock {\em Journal of the Optical Society of America A}, 19(7):1334--1345,
  2002.

\bibitem{Fienup2012}
James~R Fienup.
\newblock {Phase retrieval algorithms: a personal tour [Invited]}.
\newblock {\em Applied Optics}, 52(1):45--56, 2012.

\bibitem{Sayre1998}
D.~Sayre, H.~N. Chapman, and J~Miao.
\newblock {On the Extendibility of X-ray Crystallography to Noncrystals}.
\newblock {\em Acta Crystallographica}, A54:232--239, 1998.

\bibitem{Miao1998}
J.~Miao, D.~Sayre, and H.~N. Chapman.
\newblock {Phase retrieval from the magnitude of the Fourier transforms of
  nonperiodic objects}.
\newblock {\em Journal of the Optical Society of America A}, 15(6):1662, jun
  1998.

\bibitem{Miao1999}
Jlanwei Miao, Pambos Charalambous, Janos Kirz, and David Sayre.
\newblock {Extending the methodology of X-ray crystallography to allow imaging
  of micrometre-sized non-crystalline specimens}.
\newblock {\em Nature}, 400(6742):342--344, jul 1999.

\bibitem{Miao2012}
Jianwei Miao, Richard~L. Sandberg, and Changyong Song.
\newblock {Coherent x-ray diffraction imaging}.
\newblock {\em IEEE Journal on Selected Topics in Quantum Electronics},
  18(1):399--410, 2012.

\bibitem{Chapman2006}
Henry~N Chapman, Anton Barty, Stefano Marchesini, Aleksandr Noy, Stefan~P
  Hau-Riege, Congwu Cui, Malcolm~R Howells, Rachel Rosen, Haifeng He, John C~H
  Spence, Uwe Weierstall, Tobias Beetz, Chris Jacobsen, and David Shapiro.
\newblock {High-resolution ab initio three-dimensional x-ray diffraction
  microscopy}.
\newblock {\em Journal of the Optical Society of America A}, 23(5):1179--1200,
  2006.

\bibitem{Chen2007}
Chien-Chun Chen, Jianwei Miao, C.~W. Wang, and T.~K. Lee.
\newblock {Application of optimization technique to noncrystalline x-ray
  diffraction microscopy: Guided hybrid input-output method}.
\newblock {\em Physical Review B}, 76(6):064113, aug 2007.

\bibitem{Rodriguez2013}
Jose~A. Rodriguez, Rui Xu, Chien~Chun Chen, Yunfei Zou, and Jianwei Miao.
\newblock {Oversampling smoothness: An effective algorithm for phase retrieval
  of noisy diffraction intensities}.
\newblock {\em Journal of Applied Crystallography}, 46(2):312--318, apr 2013.

\bibitem{Abbey2008}
Brian Abbey, Keith~A. Nugent, Garth~J. Williams, Jesse~N. Clark, Andrew~G.
  Peele, Mark~A. Pfeifer, Martin {De Jonge}, and Ian McNulty.
\newblock {Keyhole coherent diffractive imaging}.
\newblock {\em Nature Physics}, 4(5):394--398, 2008.

\bibitem{Gorkhover2018}
Tais Gorkhover, Anatoli Ulmer, Ken Ferguson, Max Bucher, Filipe~R.N.C. Maia,
  Johan Bielecki, Tomas Ekeberg, Max~F. Hantke, Benedikt~J. Daurer, Carl
  Nettelblad, Jakob Andreasson, Anton Barty, Petr Bruza, Sebastian Carron, Dirk
  Hasse, Jacek Krzywinski, Daniel~S.D. Larsson, Andrew Morgan, Kerstin
  M{\"{u}}hlig, Maria M{\"{u}}ller, Kenta Okamoto, Alberto Pietrini, Daniela
  Rupp, Mario Sauppe, Gijs {Van Der Schot}, Marvin Seibert, Jonas~A. Sellberg,
  Martin Svenda, Michelle Swiggers, Nicusor Timneanu, Daniel Westphal, Garth
  Williams, Alessandro Zani, Henry~N. Chapman, Gyula Faigel, Thomas
  M{\"{o}}ller, Janos Hajdu, and Christoph Bostedt.
\newblock {Femtosecond X-ray Fourier holography imaging of free-flying
  nanoparticles}.
\newblock {\em Nature Photonics}, 12(3):150--153, mar 2018.

\bibitem{Lan2014}
Ti-Yen Lan, Po-Nan Li, and Ting-Kuo Lee.
\newblock {Method to enhance the resolution of x-ray coherent diffraction
  imaging for non-crystalline bio-samples}.
\newblock {\em New Journal of Physics}, 16(3):033016, mar 2014.

\bibitem{Barmherzig2019}
David~A. Barmherzig, Ju~Sun, Po-Nan Li, T.~J. Lane, and Emmanuel~J. Cand\`es.
\newblock {Holographic phase retrieval and reference design}.
\newblock {\em Inverse Problems}, 35(9):094001, aug 2019.

\bibitem{Miao2005}
Jianwei Miao, Yoshinori Nishino, Yoshiki Kohmura, Bart Johnson, Changyong Song,
  Subhash~H. Risbud, and Tetsuya Ishikawa.
\newblock {Quantitative image reconstruction of GaN quantum dots from
  oversampled diffraction intensities alone}.
\newblock {\em Physical Review Letters}, 95(8):085503, aug 2005.

\bibitem{Ekeberg2016}
Tomas Ekeberg, Martin Svenda, M.~Marvin Seibert, Chantal Abergel, Filipe~R.N.C.
  Maia, Virginie Seltzer, Daniel~P. DePonte, Andrew Aquila, Jakob Andreasson,
  Bianca Iwan, Olof Jonsson, Daniel Westphal, Dusko Odic, Inger Andersson,
  Anton Barty, Meng Liang, Andrew~V. Martin, Lars Gumprecht, Holger
  Fleckenstein, Sa{\v{s}}a Bajt, Miriam Barthelmess, Nicola Coppola,
  Jean~Michel Claverie, N.~Duane Loh, Christoph Bostedt, John~D. Bozek, Jacek
  Krzywinski, Marc Messerschmidt, Michael~J. Bogan, Christina~Y. Hampton,
  Raymond~G. Sierra, Matthias Frank, Robert~L. Shoeman, Lukas Lomb, Lutz
  Foucar, Sascha~W. Epp, Daniel Rolles, Artem Rudenko, Robert Hartmann, Andreas
  Hartmann, Nils Kimmel, Peter Holl, Georg Weidenspointner, Benedikt Rudek,
  Benjamin Erk, Stephan Kassemeyer, Ilme Schlichting, Lothar Struder, Joachim
  Ullrich, Carlo Schmidt, Faton Krasniqi, Gunter Hauser, Christian Reich, Heike
  Soltau, Sebastian Schorb, Helmut Hirsemann, Cornelia Wunderer, Heinz
  Graafsma, Henry Chapman, and Janos Hajdu.
\newblock {Single-shot diffraction data from the Mimivirus particle using an
  X-ray free-electron laser}.
\newblock {\em Scientific Data}, 3(1):160060, aug 2016.

\bibitem{Tkachuk2007}
Andrei Tkachuk, Fred Duewer, Hongtao Cui, Michael Feser, Steve Wang, and
  Wenbing Yun.
\newblock {X-ray computed tomography in Zernike phase contrast mode at 8 keV
  with 50-nm resolution using Cu rotating anode X-ray source}.
\newblock {\em Z. Kristallogr.}, 222:650--655, 2007.

\bibitem{Mao2019}
Wendy~L. Mao, Yu~Lin, Yijin Liu, and Jin Liu.
\newblock {Applications for Nanoscale X-ray Imaging at High Pressure}.
\newblock {\em Engineering}, 5(3):479--489, jun 2019.

\bibitem{Sierra2019}
Raymond~G. Sierra, Alexander Batyuk, Zhibin Sun, Andrew Aquila, Mark~S. Hunter,
  Thomas~J. Lane, Mengning Liang, Chun~Hong Yoon, Roberto Alonso-Mori, Rebecca
  Armenta, Jean~Charles Castagna, Michael Hollenbeck, Ted~O. Osier, Matt Hayes,
  Jeff Aldrich, Robin Curtis, Jason~E. Koglin, Theodore Rendahl, Evan
  Rodriguez, Sergio Carbajo, Serge Guillet, Rob Paul, Philip Hart, Kazutaka
  Nakahara, Gabriella Carini, Hasan Demirci, E.~Han Dao, Brandon~M. Hayes,
  Yashas~P. Rao, Matthieu Chollet, Yiping Feng, Franklin~D. Fuller, Christopher
  Kupitz, Takahiro Sato, Matthew~H. Seaberg, Sanghoon Song, Tim~B. {Van Driel},
  Hasan Yavas, Diling Zhu, Aina~E. Cohen, Soichi Wakatsuki, and S{\'{e}}bastien
  Boutet.
\newblock {The macromolecular femtosecond crystallography instrument at the
  linac coherent light source}.
\newblock {\em Journal of Synchrotron Radiation}, 26(2):346--357, mar 2019.

\bibitem{Xiao2009}
Chuan Xiao, Yurii~G Kuznetsov, Siyang Sun, Susan~L Hafenstein, Victor~A
  Kostyuchenko, Paul~R Chipman, Marie Suzan-Monti, Didier Raoult, Alexander
  McPherson, and Michael~G Rossmann.
\newblock {Structural Studies of the Giant Mimivirus}.
\newblock {\em PLoS Biology}, 7(4):e1000092, apr 2009.

\bibitem{Miao2003}
Jianwei Miao, Tetsuya Ishikawa, Erik~H. Anderson, and Keith~O. Hodgson.
\newblock {Phase retrieval of diffraction patterns from noncrystalline samples
  using the oversampling method}.
\newblock {\em Physical Review B}, 67(17):174104, may 2003.

\bibitem{DeCaro2010}
Liberato {De Caro}, Elvio Carlino, Gianvito Caputo, Pantaleo~Davide Cozzoli,
  and Cinzia Giannini.
\newblock {Electron diffractive imaging of oxygen atoms in nanocrystals at
  sub-ngstr{\"{o}}m resolution}.
\newblock {\em Nature Nanotechnology}, 5(5):360--365, apr 2010.

\end{thebibliography}

\end{document}